\documentclass[amsmath,amssymb,a4paper, superscriptaddress]{revtex4}
\usepackage{graphicx}
\usepackage{amsmath}

\begin{document}
\title{Quasinormal Modes in three-dimensional time-dependent Anti-de Sitter spacetime}
\author{Zai-Xiong Shen}
\author{Bin Wang}
\affiliation{Department of Physics, Fudan University, 200433
Shanghai, People's Republic of China}
\author{Ru-Keng Su}
\affiliation{Chinese Center of Advanced Science and Technology
(World Lab), P. O. Box 8730, 100080 Beijing, People's Republic of
China}

\begin{abstract}
The massless scalar wave propagation in the time-dependent BTZ
black hole background has been studied. It is shown that in the
quasi-normal ringing both the decay and oscillation time-scales
are modified in the time-dependent background.
\end{abstract}

\maketitle

\makeatletter
\def\@captype{figure}
\makeatother

\setlength{\parskip}{5pt}

Quasinormal modes of black holes have been an intriguing subject
of discussions over thirty years \cite{1}, leading to important
contributions to the understanding of black holes \cite{2,3,4,5}.
Up until very recently, all works in this field deal with
asymptotically flat spacetimes. In the past few years the study
has been extended to de Sitter space \cite{6,7} as well as
Andi-de-Sitter space \cite{8,9,10,11}. In addition to the
astronomical interest that quasinormal modes carry a unique
fingerprint which would lead to the direct identification of the
black hole existence, quasinormal modes is a good testing ground
which gives evidence of the correspondence between gravity in
AdS(dS) spacetime and quantum field theory at the boundary.

So far the study of quasinormal modes is restricted to
time-independent black hole backgrounds. It should be realized
that it is realistic to consider a black hole with parameters
changing with time due to the absorption or evaporation processes.
Recently the influence of the time-dependent spacetime effect on
the late-time tail behavior of the perturbation on the black hole
background has been explored \cite{12}. In our previous work
\cite{13}, we have investigated the modification to the
quasinormal modes in the dynamic Schwarzschild black hole
background. The temporal evolution of massless scalar field
perturbation, especially the quasinormal modes in different
time-dependent situation have been obtained. It is of interest to
extend our study to time-dependent AdS spacetime. AdS spacetime
share with asymptotically flat spacetimes the common property
which makes it a good testing ground what one wants to go beyond
asymptotic flatness. Besides exploring the quasinormal modes in
time-dependent AdS spacetime can provide further understanding to
the AdS/CFT correspondence.

We will concentrate on 3D non-rotating BTZ black hole \cite{14} with the metric
\begin{equation}
ds^2=\left(-M+\frac{r^2}{l^2}\right)dt^2-\left(-M+\frac{r^2}{l^2}\right)^{-1}dr^2-r^2d\phi^2
\label{eq:1}
\end{equation}
and $\Lambda=-1/l^2$

In this general stationary coordinate, the quasinormal modes and
associated frequencies are studied in \cite{11}. However, this
coordinate is not appropriate to be used to study the
time-dependent case \cite{13}. One option is to use the
Kruskal-like coordinate in time-dependent problems. Here we first
show that Kruskal coordinate is valid in studying the wave
propagation in stationary BTZ black hole.

The Kruskal coordinate for the BTZ black hole is given by
\cite{14}
\begin{equation}
ds^2=\Omega ^2\left( d\tau ^2-d\rho ^2\right) -r^2d\phi^2
\label{eq:2}
\end{equation}
where $\tau$ and $\rho$ are the time-like and space-like
coordinates respectively. For $r \ge r_+$, they are defined by

\begin{equation}
r_+<r\leq \infty \left \{
    \begin{array}{lcl}
        \rho & = & \sqrt{\frac{r-r_+}{r+r_+}}\cosh a_+t
        \\[20pt]
        \tau & = & \sqrt{\frac{r-r_+}{r+r_+}}\sinh a_+t \\
    \end{array}
\right.
\label{eq:3}
\end{equation}
where
\[
    a_+=\frac{\sqrt{M}}{l},\;\;\;\;\;
    r_+=l\sqrt{M},\;\;\;\;\;
    \Omega^2(r,t)=\left( \frac r{\sqrt{M}}+l \right) ^2
\]

The relation between $\tau$, $\rho$ and $t$, $r$ are

\begin{equation}
    t=\frac{l}{2\sqrt{M}}\ln \left( \frac{1+\tau /\rho}{1-\tau
    /\rho} \right)
    \label{eq:4}
\end{equation}

\begin{equation}
    r=\frac{1+(\rho ^2 - \tau ^2)}{1-(\rho ^2 - \tau ^2)}l\sqrt{M}
    \label{eq:5}
\end{equation}

The propagation of the massless scalar field in Kruskal
coordinates is governed by

\begin{equation}
    \frac 1{\sqrt{r}}\left( \Psi _{,\,\tau \tau }-\Psi _{,\,\rho \rho }\right)
    +\left( -\frac{r'^2-\dot r^2}{4r^{5/2}}+\frac{r''-\ddot{r}}
    {2r^{3/2}}+\frac{m^2}{r^{5/2}}\Omega ^2\right) \Psi=0
\label{eq:6}
\end{equation}
where $m$ is the angular quantum number.

Analogous to the null coordinates used in \cite{4}, we make the variable transformations
\begin{equation}
    u = \tau -\rho, \;\;\;\;\; v = \tau + \rho
    \label{eq:7}
\end{equation}
then the expression of $t$ and $r$ are
\begin{equation}
    t=\frac{l}{2\sqrt{M}}\ln \left( -\frac vu\right)
    \label{eq:8}
\end{equation}

\begin{equation}
    r=\frac{1-uv}{1+uv}l\sqrt{M}
    \label{eq:9}
\end{equation}
and $r$ will be a function of $u$ and $v$, $r=r(u,v)$, and
$\frac{\partial }{\partial \tau }=\frac{\partial }{\partial
v}+\frac{\partial }{\partial u}$, $\frac{\partial }{\partial
\rho }=\frac{\partial }{\partial v}-\frac{\partial }{\partial
u}$.

Equation (\ref{eq:6}) becomes
\begin{equation}
    \frac 4{\sqrt{r}}\Psi _{,\,uv}+
    \left( \frac 1{r^{5/2}}\frac{\partial r}{\partial v}\frac{\partial r}{\partial u}
    -\frac 2{r^{3/2}}\frac{\partial ^2r}{\partial u\partial v}+\frac{m^2}{r^{5/2}}\Omega ^2\right) \Psi=0
    \label{eq:10}
\end{equation}
Taking
\[
    u^*=\ln\left( -\frac{1}{u}\right)
    \;\;\;\;\;\;\;\;\;\;\;\;\;\;\;\;\;\;\;\;\;\;\;\;\;\;\;\;\;\;\;\;\;\;\;
    v^*=\ln v
\]

we have

\begin{equation}
    \left \{
        \begin{array}{lcl}
            t&=&\frac{l}{\sqrt{M}}\frac {v^*+u^*}{2}\\[20pt]
            r^*&=&\frac{l}{\sqrt{M}}\frac {v^*-u^*}{2}\\[20pt]
            r&=&-l\sqrt{M}\coth \left( \frac{\sqrt M}{l}r_*\right)
        \end{array}
    \right.
    \label{eq:11}
\end{equation}

Eq.(\ref{eq:10}) can be written as
\begin{equation}
    -4\Psi _{,\,u^*v^*}=V\Psi
    \label{eq:12}
\end{equation}
where the effective potential reads
\begin{equation}
    V=\frac{r-r_+}{r+r_+}\left( \frac 1{r^2}\frac{\partial r}{\partial v}\frac{\partial r}{\partial u}
    -\frac 2r\frac{\partial ^2r}{\partial u\partial v}+\frac{m^2}{r^2}\Omega ^2\right)
    \label{eq:13}
\end{equation}

This potential diverges when $r\rightarrow\infty$, which has the
same property as that in general coordinates (1). This
divergence can be overcome by imposing the boundary condition
$\Phi =0$ as $r\rightarrow\infty$($u^*=v^*$).

It is straightforward to write Eq. (\ref{eq:12}) into the discrete
form
\begin{equation}
    \Psi_N = \Psi_W + \Psi_E - \Psi_S -\Delta u^* \Delta v^*
    V\frac{\Psi_W + \Psi_E}{8}
    \label{eq:14}
\end{equation}

As one can see from the TABLE. \ref{table:1}, numerical result got
by employing the Kruskal coordinates agrees well to that of the
general coordinate \cite{11}. This shows that Kruskal coordinate
is valid to investigate the wave propagation in the BTZ black hole
background.

From the wave equation in the Kruskal coordinate (\ref{eq:6}), it
is easy to see that the metric terms of the time-like coordinate
$\tau$ and the space-like coordinate $\rho$ are the same and black
hole parameters only appear in the effective potential. This makes
the investigation in the time-dependent case much easier. In the
time-dependent case, the wave equation has the same form
(\ref{eq:12}), however now
\begin{eqnarray}
    \frac{\partial ^2r}{\partial u\partial v}&=&\frac 1{1+uv}\left[ \frac{l(1-uv)}{2\sqrt{M}}\left(
    M_{,uv}-\frac{M_{,u}M_{,v}}{2M}\right) \right. \nonumber \\
    &&\left. -\frac l{(1+uv)\sqrt{M}}\left( vM_{,v}+uM_{,u}\right) -\frac{2(1-uv)}{(1+uv)^2}l\sqrt{M} \right] \nonumber \\
    &=&\frac{r}{2M}\left( M_{,uv}-\frac{M_{,u}M_{,v}}{2M}\right)
    -\frac{(r+r_+)^2}{4r_+M}\left( vM_{,v}+uM_{,u}\right) \nonumber \\
    &&-\frac r2\left( 1+\frac{r}{r_+}\right) ^2
    \label{eq:15}
\end{eqnarray}
and
\begin{eqnarray}
    \frac{\partial r}{\partial u}\frac{\partial r}{\partial v}&=&\frac{l^2}{(1+uv)^2}\left[
    \frac{4uv}{(1+uv)^2}M \right. \nonumber \\
    &&\left. +\frac{(1-uv)^2}{4}\frac{M_{,u}M_{,v}}{M}-\frac{1-uv}{1+uv}\left( vM_{,v}+uM_{,u}\right) \right] \nonumber \\
    &=&\frac{r^2}{4M^2}M_{,u}M_{,v}-\frac{r(r+r_+)^2}{4r_+M}\left( vM_{,v}+uM_{,u}\right) \nonumber \\
    &&-\frac{(r+r_+)^3(r-r_+)}{4r_+^2}
    \label{eq:16}
\end{eqnarray}
where the mass of the black hole has a general dependence on $t$,
then it is a function of $\tau$ and $\rho$ in Kruskal coordinate.

We now present the result of our numerical calculations in the
time-dependent black hole background. In the first series of
simulations, we consider the simple situation by choosing the mass
of the black hole $M=M_0+\frac{2A}{l}t$, where $M_0,A$ are
constants. Employing (\ref{eq:8}), we have
\begin{equation}
    M^{3/2}-M_0M^{1/2}=A\ln\left( -\frac{v}{u}\right)
    \label{eq:17}
\end{equation}

and

\[
    M_{,u}=-\frac{2A\sqrt{M}}{(3M-2M_0)u}
    \;\;\;\;\;\;\;\;\;\;\;\;\;\;\;\;\;\;\;
    M_{,v}=\frac{2A\sqrt{M}}{(3M-2M_0)v}
\]

\begin{equation}
    \left \{
        \begin{array}{lcl}
            M_{,u}M_{,v}&=&\frac{4A^2M\left( r+r_+\right)}
            {9(3M-2M_0)^2\left( r-r_+\right)}\\[20pt]
            M_{,uv}&=&-\frac{3M+2M_0}{2M(3M-2M_0)}M_{,u}M_{,v}
        \end{array}
    \right.
    \label{eq:18}
\end{equation}

\begin{figure}
\includegraphics{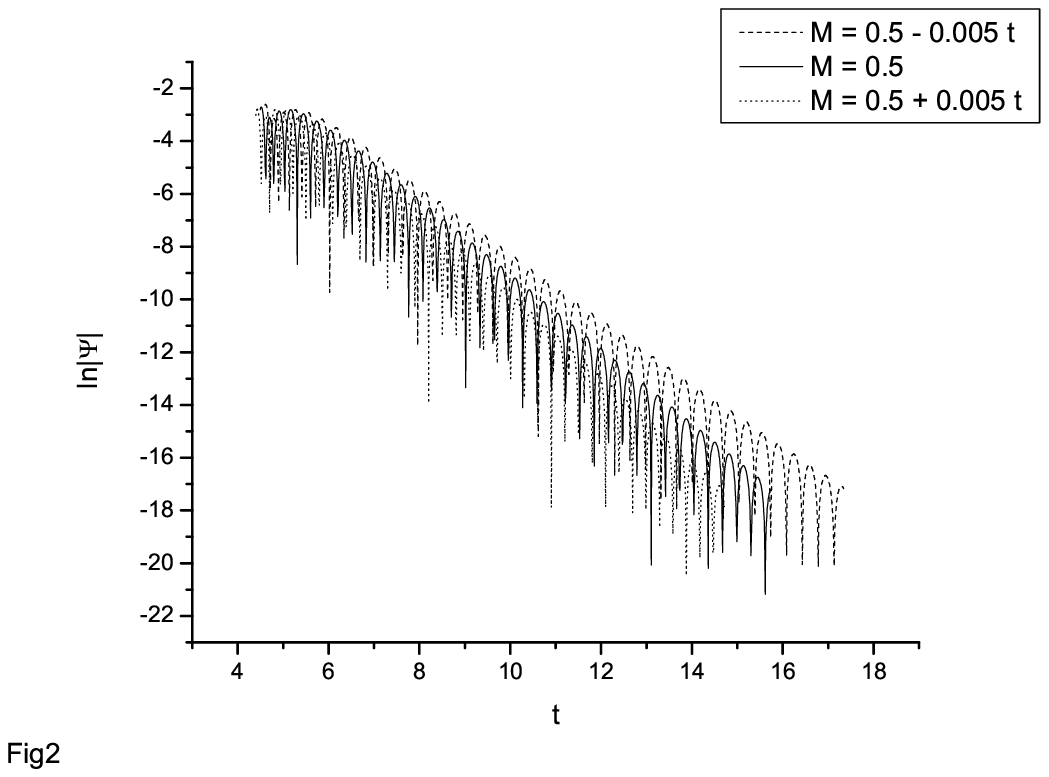}
\caption{\label{fig:2}Temporal evolution of the field in the
background of a BTZ black hole for $m = 10$. The mass of
the black hole $M(t) = M_0 \pm 2At$, where $M_0 = 0.5$ and $2A = 5
\times 10^{-3}$ are constant coefficients. The field evolution for
$M = M_0 + 2At$ and $M = M_0 - 2At$ are shown as the bottom curve
and the top curve respectively. For comparison, the oscillation
for $M = M_0$ is also displayed (the middle curve).}
\end{figure}

\begin{figure}
\includegraphics{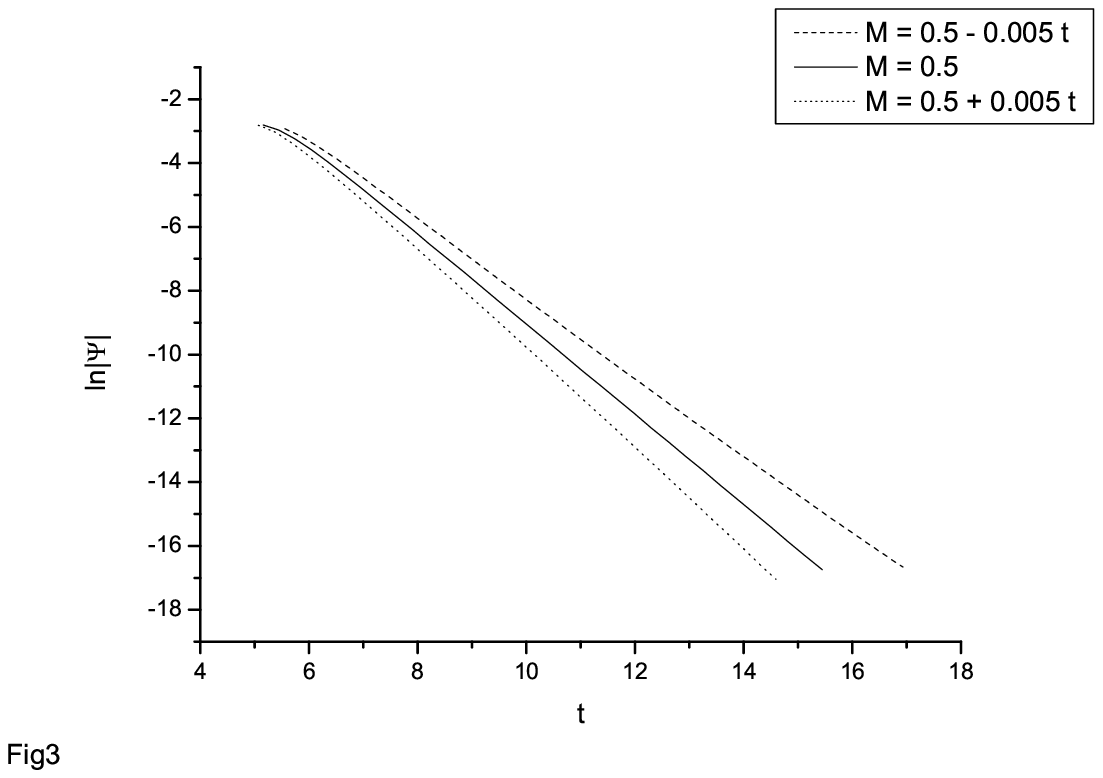}
\caption{\label{fig:3}The same plot as Fig. \ref{fig:2}. For
clarity, the connected maxima of the oscillations are displayed.
The imaginary frequency $\omega_I$ can be read from the decay time
scale $\tau\sim 1/\omega_I$.}
\end{figure}

\begin{figure}
\includegraphics{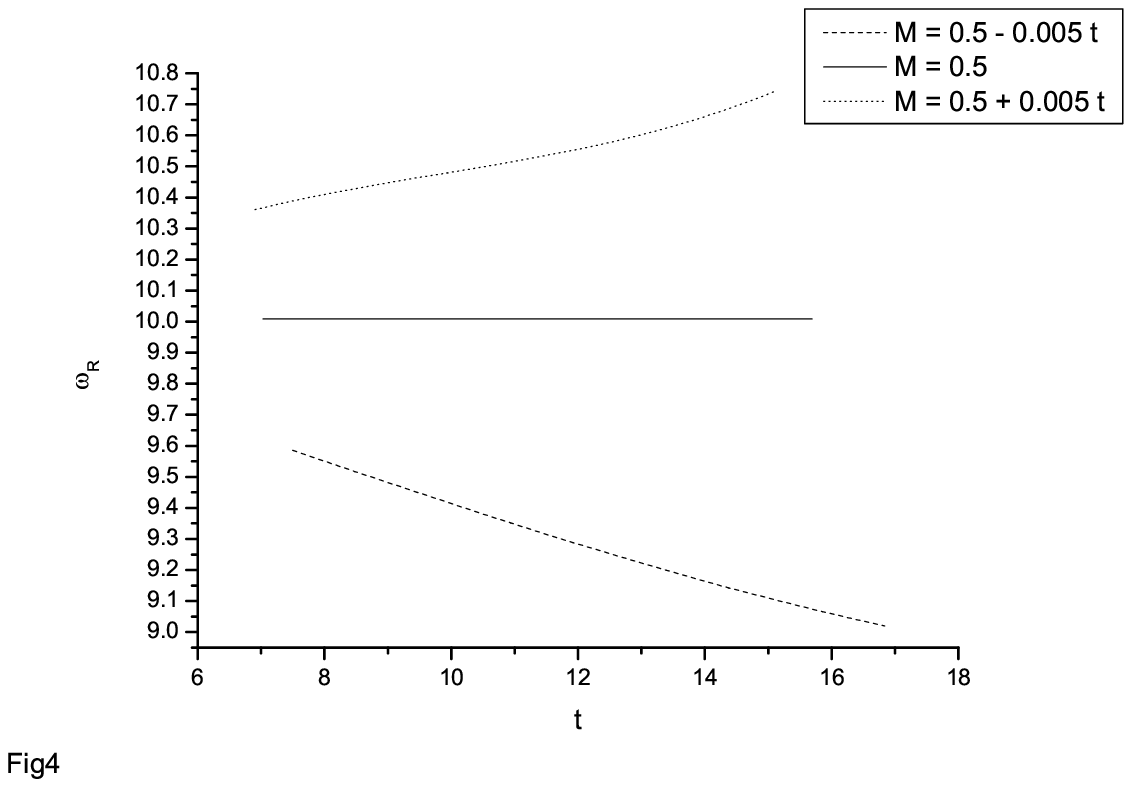}
\caption{\label{fig:4}The frequency $\omega_R$ determined from
the oscillation period in Fig. \ref{fig:2} is shown as a function
of $t$.}
\end{figure}

The results are shown in Fig. \ref{fig:2}, \ref{fig:3} and
\ref{fig:4}. The modification to the quasinormal modes due to the
time-dependent background is clear. When $M$ increase linearly
with $t$, the decay becomes faster compared to the stationary
case, which corresponds to say that $\omega_I$ increases with $t$.
The real part of the quasinormal frequency is no longer a constant
as that for stationary black hole, it increases with the increase
of time. When $M$ decreases linearly with $t$, compared to the
stationary case, we observed that both $\omega_I,\omega_R$ decrease
with the increase of time. The situation of the quasinormal
frequency on the evolution of time is different from that in
asymptotically flat spacetime \cite{13}. This difference is caused by the
special property of AdS spacetime which leads to different
behavior of the effective potential from that of the
asymptotically flat spacetime. In light of the observations of
Ching et al \cite{15} this different result is not surprising.

We have also extended our discussion to a more realistic model, an
evaporating black hole with the mass determined by \cite{16}
\begin{equation}
    \dot{M} = -\alpha_0M^2
    \label{eq:19}
\end{equation}
where $\alpha_0$ is a constant coefficient. One can obtain the
mass of the black hole as a function of $t$ from Eq. (\ref{eq:19})
as
\begin{equation}
    M(t)=\frac{1}{\alpha_0t+b}
\end{equation}
where $b$ is an arbitrary constant.

\begin{figure}
\includegraphics{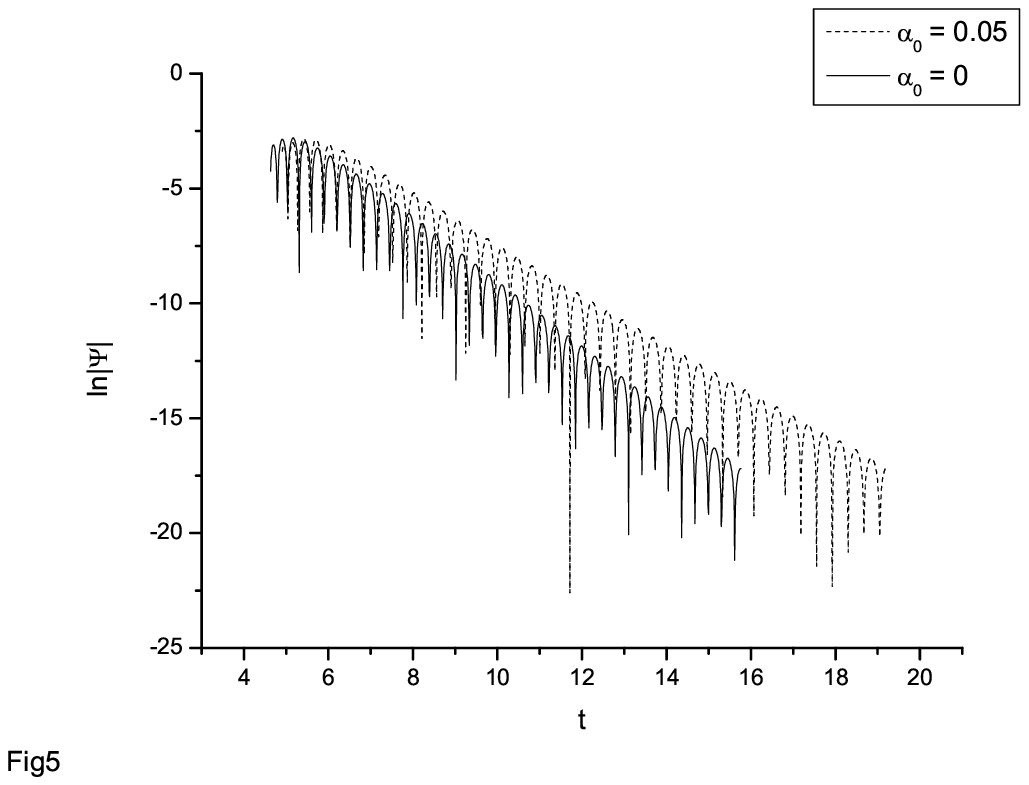}
\caption{\label{fig:5}Temporal evolution of the field in the
background of a evaporating BTZ black hole (upper curve). We use
$\alpha_0 = 0.05$ and $b = 2.0$ in our calculation. For
comparison, the mode in the stationary background with $M = 0.5$
($\alpha_0 = 0$ and $b = 2.0$) is also displayed (lower curve).}
\end{figure}

\begin{figure}
\includegraphics{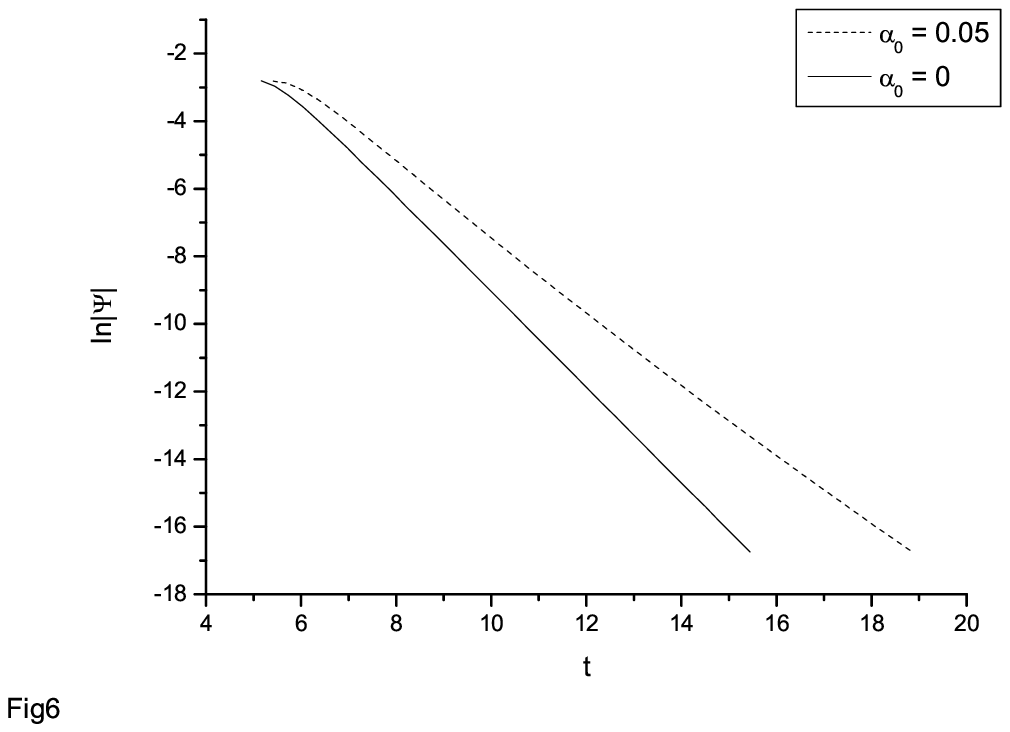}
\caption{\label{fig:6}Same as Fig. \ref{fig:5}. Only the connected
maxima of the oscillations are displayed. The imaginary frequency
$\omega_I$ can be read from the decay time scale $\tau\sim
1/\omega_I$.}
\end{figure}

\begin{figure}
\includegraphics{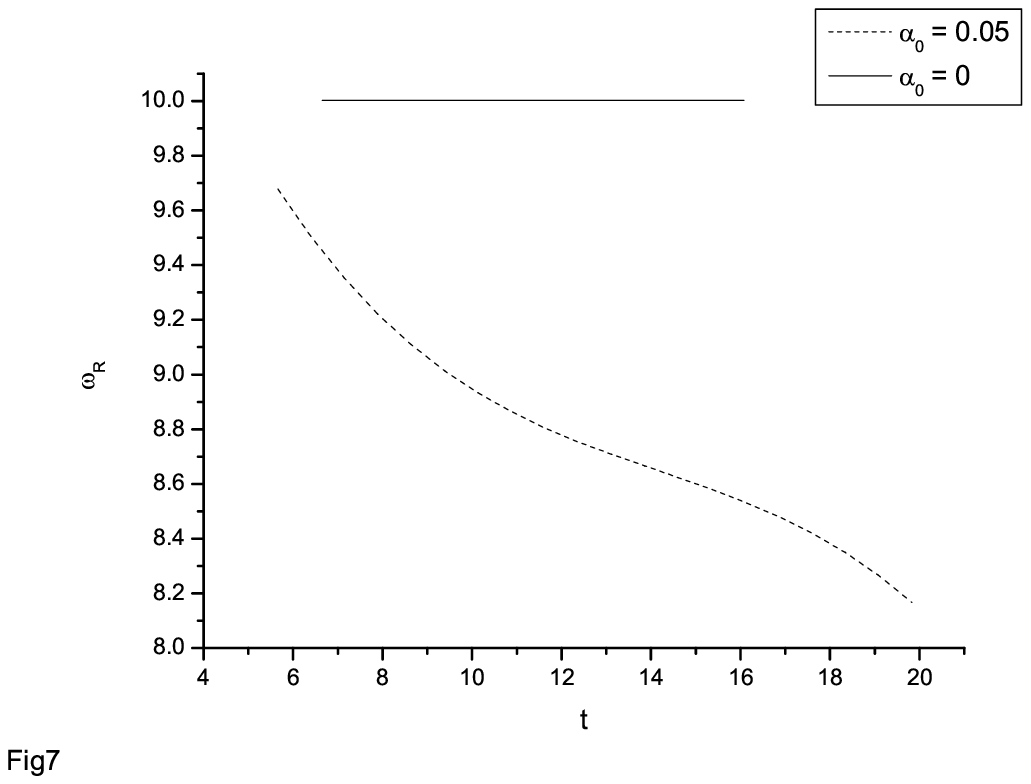}
\caption{\label{fig:7}The frequency $\omega_R$ determined from
the oscillation period in Fig. \ref{fig:5} is shown as a function
of $t$ for the evaporating black hole. For the stationary black
hole $\omega_R$ is a constant.}
\end{figure}

Results of the numerical calculations are shown in Fig.
\ref{fig:5}, \ref{fig:6}, and \ref{fig:7}. Different from the
stationary black hole case, for the evaporating black hole both
real and imaginary parts of quasi-normal frequencies $\omega_I$
and $\omega_R$ decrease with respect to $t$, in consistent with
the simple case above.

In summary we have studied the evolution of the massless scalar
field in the time-dependent BTZ black hole background. We have
found that the Kruskal-like coordinate is an appropriate framework
to investigate the wave propagation in the time-dependent
spacetimes.  In our study we
have tried to derive the time-dependent potential in a natural way
by considering dynamic black holes with black hole parameters
changing with time. In our numerical study, we have found the
modification to the QNM due to the temporal dependence of the
black hole spacetimes. The decay and oscillation timescale are no
longer constants with the evolution of time as that in the
stationary black hole case. In the absorption process, when the
black hole mass becomes bigger, both the real and imaginary parts
of the quasi-normal frequencies increase with the increase of $t$.
However, in the evaporating process, when the black hole loses
mass, both the real and imaginary parts of the quasi-normal
frequencies decrease with the increase of $t$. This property is
different from that of the asymptotically flat spacetime \cite{13} and
is caused by the special characteristic of behavior of the
effective potential in AdS spacetime. Our result consolidate the
argument proposed in \cite{15} in time-dependent situation that
effective potential influences a lot on the quasinormal modes.

This work was  partially supported by National Natural Science
Foundation of China under  grant 10005004, 10047005, 10247001,
10235030 and the foundation of Ministry of Education of China and
Shanghai Science and Technology Commission.

\begin{table}[!htp]
\tabcolsep 20pt \caption{Quasi-normal frequencies for the scalar
perturbation in the stationary BTZ background.
\label{tab:1}}

\begin{tabular}{c|c||c|c|c|c}
\hline \hline &&\multicolumn{2}{c|}
{\textbf{General Coordinate}}&\multicolumn{2}{c}{\textbf{Kruskal Coordinate}} \\
\hline
$\sqrt{M}$&$m$&$\omega_R$&$\omega_I$&$\omega_R$&$\omega_I$ \\
\hline
0.2&1&1&-0.4&1.002&-0.400 \\
0.4&1&1&-0.8&0.965&-0.799 \\
0.4&2&2&-0.8&2.003&-0.800 \\
0.4&3&3&-0.8&3.004&-0.800 \\
0.4&4&4&-0.8&4.005&-0.800 \\
0.5&2&2&-1&1.999&-1.000 \\
0.5&3&3&-1&2.993&-1.000 \\
0.5&4&4&-1&4.007&-1.000 \\
1&4&4&-2&4.004&-2.001 \\
2&10&10&-4&10.014&-4.001 \\
3&10&10&-6&10.028&-6.011 \\
4&10&10&-8&9.973&-7.983 \\
\hline \hline
\end{tabular}

\label{table:1}
\end{table}

\begin{figure}
\includegraphics{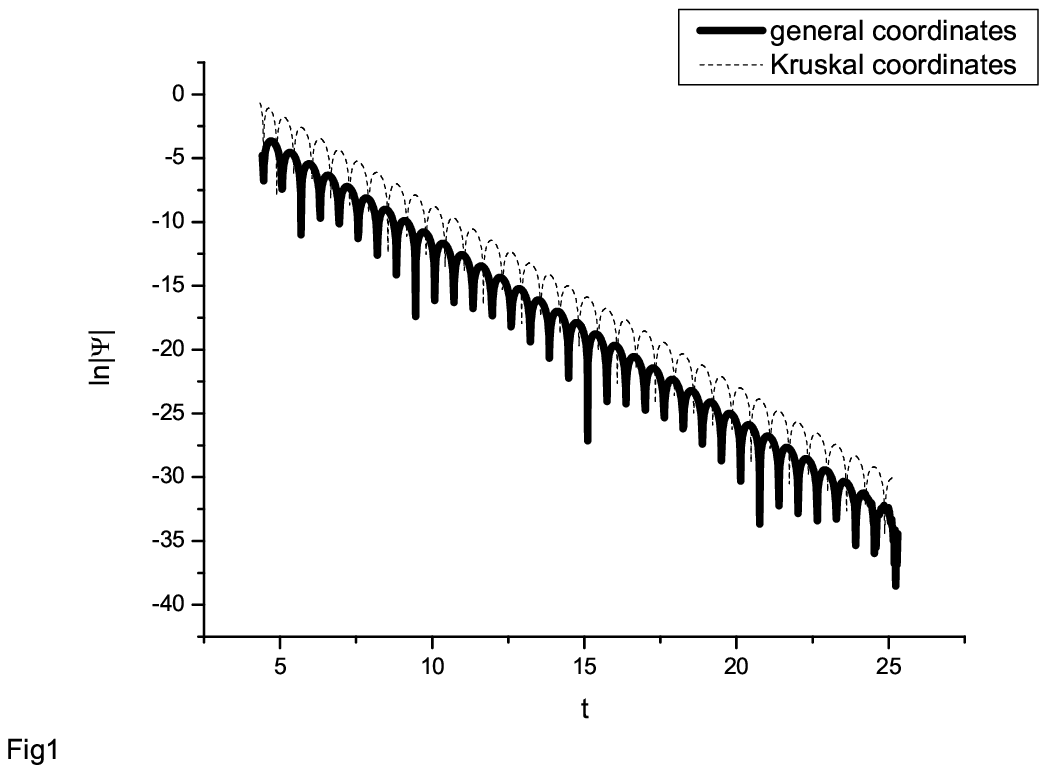}
\caption{\label{fig:1}Temporal evolution of the field $\Psi$ at a
fixed radius. The mass of the black hole $M = 0.5$ and the
multipole index $m=5$.}
\end{figure}

\begin{thebibliography}{99}
\bibitem{1} K. D. Kokkotas and B. G. Schmidt, Living Rev. Rel. \textbf{2}, (1999); H. P. Nollert, Class. Quant. Grav. \textbf{16}, R159 (1999).
\bibitem{2} K. D. Kokkotas and B. F. Schwtz, Phys. Rev. D \textbf{37}, 3378 (1988); N. Andersson, Proc. R. Soc. London A \textbf{442}, 427 (1993); E. W. Leaver, Phys. Rev. D \textbf{41}, 2986 (1990).
\bibitem{3} R. H. Price, Phys. Rev. D \textbf{5}, 2419 (1972)
\bibitem{4} C. Gundlach, R. H. Price, and J. Pullin, Phys. Rev. D \textbf{49}, 883 (1994)
\bibitem{5} S. Hod and T. Piran, Phys. Rev. D \textbf{58}, 044018 (1998); R. Moderski and M. Rogatko, Phys. Rev. D \textbf{64}, 044024 (2001); H. Koyama and A. Tomimatsu, Phys. Rev. D \textbf{63}, 064032 (2001); L. H. Xue, B. Wang and R. K. Su, Phys. Rev. D \textbf{66}, 024032 (2002).
\bibitem{6} P. R. Brady, C. M. Chambers, W. Krivan and P. Lagunas, Phys. Rev. D \textbf{55}, 7538 (1997); P. R. Brady, C. M. Chambers, W. G. Laarakkers and E. Poisson, Phys. Rev. D \textbf{60}, 064003 (1999).
\bibitem{7} E. Abdalla, B. Wang, A. Lima-Santos and W. G. Qiu, Phys. Lett. B \textbf{538}, 435 (2002); E. Abdalla, K. H. C. Castello-Branco and A. Lima-Santos, Phys. Rev. D \textbf{66}, 104018 (2002).
\bibitem{8} J. S. F. Chan and R. B. Mann, Phys. Rev. D \textbf{55}, 7546 (1999); ibid \textbf{59}, 064025 (1999).
\bibitem{9} G. T. Horowitz and V. E. Hubney, Phys. Rev. D \textbf{62}, 024027 (2000); G. T. Horowitz, Class. Quant. Grav. \textbf{17}, 1107 (2000).
\bibitem{10} B. Wang, C. Y. Lin and E. Abdalla, Phys. Lett. B \textbf{481}, 79 (2000); B. Wang, C. Molina and E. Abdalla, Phys. Rev. D \textbf{63}, 084001 (2001); J. M. Zhu, B. Wang and E. Abdalla, Phys. Rev. D \textbf{63}, 124004 (2001); B. Wang, E. Abdalla and R. B. Mann, Phys. Rev. D \textbf{65}, 084006 (2002).
\bibitem{11} V. Cardoso and J. P. Lemos, Phys. Rev. D \textbf{63}, 124015 (2001).
\bibitem{12} S. Hod, Phys. Rev. D \textbf{66}, 024001 (2002).
\bibitem{13} L. H. Xue, Z. X. Shen, B. Wang, R. K. Su, gr-qc/0304109
\bibitem{14} M. Banados, C. Teitelboim and J. Zanelli, Phy. Rev. Lett \textbf{69}, 1849 (1992), M. Banados, M. Henneaux, C. Teitelboim, J. Zanelli, Phy. Rev. D \textbf{48}, 1506 (1993); S. Carlip, Class. Quant. 2853 (1995)
\bibitem{15} E. S. C. Ching, P. T. Leung, W. M. Suen and K. Young, Phy. Rev. D \textbf{52}, 2118 (1995); Phy. Rev. Lett \textbf{74}, 2414 (1995).
\bibitem{16} B. Wang, J. M. Zhu, Mod. Phys. Lett. A \textbf{12}, 1298 (1995)
\bibitem{17} D. Page, Phys. Rev. D \textbf{13}, 198 (1976); W. A. Hiscock and L. D. Weems, Phys. Rev. D \textbf{41}, 1142 (1990).

\end{thebibliography}
\end{document}